\documentclass[aps,showpacs,eps,preprint]{revtex4}
\usepackage{colordvi}
\usepackage{amssymb}
\usepackage{amsmath}
\usepackage{epsf}
\usepackage{mathrsfs}
\usepackage[dvips]{graphics}
\usepackage[dvips,usenames]{color}
\begin{document}
%\draft
\def \beq{\begin{equation}}
\def \eeq{\end{equation}}
\def \bea{\begin{eqnarray}}
\def \eea{\end{eqnarray}}
\def \bem{\begin{displaymath}}
\def \eem{\end{displaymath}}
\def \P{\Psi}
\def \Pd{|\Psi(\boldsymbol{r})|}
\def \Pds{|\Psi^{\ast}(\boldsymbol{r})|}
\def \Po{\overline{\Psi}}
\def \bs{\boldsymbol}
\def \bl{\bar{\boldsymbol{l}}}
\title{ Cold atoms in rotating optical lattice with nearest neighbour interaction}
\author{ Rashi Sachdeva, Sonika Johri and Sankalpa Ghosh }
\affiliation{Department of Physics, Indian Institute of Technology, Delhi, New Delhi-110016}
\begin{abstract}
Extended Bose Hubbard models with nearest neighbour interaction describe minimally the effect of long range interaction
on ultra cold atoms in deep optical lattices. Rotation of such optical lattices subject such neutral cold atoms
to the effect of an  artificial magnetic field. The modification of the phase boundaries of the density wave and Mott Insulator phases due to this
rotation are shown to be related to the edge spectrum of spinorial and scalar Harper equation. Corresponding profiles of the checkerboard vortex states with sublattice modulated superfluid order parameter near density wave phase boundary are calculated.
\end{abstract}
\date{\today}
\maketitle

\section{Introduction}
Ultra cold atomic condensates
with short range interaction in deep optical lattices are described by the Bose Hubbard model \cite{ Jaksch} in the tight binding
approximation and shows quantum phase transition from the superfluid (SF) to Mott insulator (MI) phase due to the competition between nearest neighbour hopping and on site interaction \cite{Greiner}. Such condensates when subjected to an artificial magnetic field through rotation \cite{vortices} or by imprinting motion dependent laser induced phases on their internal states \cite{synthetic}, form vortices.  The effect of an artificial magnetic field on the phases of cold atoms generated in the presence of an optical lattice \cite{Niemeyer} can either by studied by trapping more than one internal states of the atom in optical lattice \cite{ Zoller, Gerbier}
or by rotating the optical lattice \cite{raprot}. This explores the effect of an artificial magnetic field on ultra cold neutral atoms in tight binding approximation.

\begin{figure}[ht]
\centerline{ \epsfxsize 8cm \epsfysize 4cm \epsffile{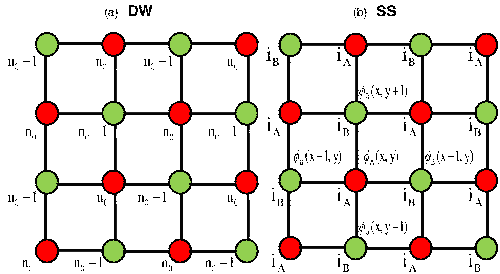}}
\caption{{(a) alternating particle number in density wave phase (b) superfluid order parameter in super solid phase on the sites of A (red) and
B(green) sublattices.}}
\label{fig_latt}
\end{figure}

Extended Bose Hubbard (EBH) model \cite{White} that includes additionally interaction between atoms at different lattice sites described
such cold atoms in optical lattices with long range interaction \cite{Santos}.
Examples are dipolar cold atoms or polar molecules \cite{dipolar}. In this paper we study the effect of rotation on such
EBH model that includes nearest neighbour interaction (NNI) apart from the on-site interaction. The
addition of the NNI to the Bose-Hubbard hamiltonian has pronounced effect on the phases since
the corresponding phase diagram \cite{ Kovrizhin, Bartouni, Pai} contains the density wave (DW) and supersolid (SS) phases apart from
the MI and SF phases. The DW and MI phases lack coherence as the SF order parameter vanishes. Both have fixed number of particles at a given site. But DW has alternating particle numbers on successive sites ( Fig. \ref{fig_latt} (a)) where as in the MI phase they are uniform.

In the intriguing supersolid (SS)  phase the superfluid order parameter and the crystal order co-exist and the superfluid density gets spatially modulated.
The supersolid phase was first experimentally cited in solid helium \cite{Mose} though the interpretation of the experimental results
was not without controversy \cite{Reppy}. However if realized with cold atomic system in optical lattice, such a supersolid phase
can be identified in a clearer fashion. An unambiguous way of identifying the SS phase is to study the modulation
of the superfluid order in the vortices created in such phase which will be different from the vortices created in an uniformly rotated superfluid.
To understand such vortex profiles one thus need to study the effect of such gauge field on the phases of EBH model.

The phase diagram of ordinary BH model in presence of such gauge field or equivalently cold atoms in rotated optical lattice
recently inspired a number of work \cite{pinning, Bhat, Zhang, Oktel, Mueller1, Lundh, Scarola, Sengupta, Powell}. The change in the
nature of the quasiparticle excitations both near the phase boundary \cite{Zhang, Oktel, Mueller1, Sengupta} as well as deep inside the superfluid phase due to the effect of the gauge field \cite{Powell} has been studied extensively. However, the effect of  gauge field on SS phases realized in EBH model received much less attention. In this paper we report  the modification of the DW-SS phase boundary and
the novel vortex profiles in SS phase near such phase boundary due to such gauge field.

We calculate the modification of the DW phase boundary in the mean field approximation by using a
reduced basis ansatz for the Gutzwiller variational wavefunction. The minimization of the energy functional very close to the DW phase boundary
shows that the superfluid order parameter satisfies a spinorial Harper equation \cite{Harper}. Consequently the phase boundary can be determined from the edge of a Hofstadter butterfly (HB) spectrum \cite{Hofstadter}. In the resulting vortices, the spatial profile of the superfluid density shows
a checkerboard like two sublattice modulation with a relative phase winding between the superfluid order parameter defined on each of these sublattices.
We discuss their possible experimental detection.

\section{Theoretical Framework}
We consider a square optical lattice in two spatial dimension rotated in the plane about $z$ axis. The corresponding
tight binding Hamiltonian in the co-rotating frame with onsite interaction and NNI is  given by
\bea  \hat{H} & = & -t \sum_{\langle i, j
\rangle}(\hat{a}_{i}^{\dag}\hat{a}_{j}\exp(i \varphi_{ij}) + \text{\it h.c.}) \nonumber \\
& & \mbox{} +\frac{1}{2}\sum_{i}\hat{n}_{i}(\hat{n}_{i}-1)
+ V\sum_{\langle i,j \rangle}\hat{n}_{i}\hat{n}_{j}
-\mu\sum_{i}\hat{n}_{i}
\label{RBH} \eea
Here hopping amplitude $t$, NNI strength $V$ and chemical potential $\mu$ are expressed in unit of the on site repulsion energy $U$. $\langle \rangle$ implies that site index $i,j$ on the two dimensional square lattice are the nearest neighbours and $\hat{a}_{i}^{\dag},\hat{a}_{i},\text{and}~ \hat{n}_{i}$ are bosonic creation, annihilation and number operators for the $i$-th site. We neglect the effect
of an overall trap potential assuming that it is sufficiently shallow and gets neutralized by the centrifugal force that normally happens in the bulk of the system.

The phase factor $\varphi_{ij}=\int_{r_{j}}^{r_{i}}d\bs{r}.\bs{A}(r)$ with the
effective vector potential $\bs{A}(r)=(m/\hbar)(\bs{\Omega} \times  \bs{r})= \pi\nu(x\hat{y}-y\hat{x})$ in the symmetric gauge.
The resulting artificial magnetic field is $2 \Omega \hat{z}$  where $\Omega$ is the frequency of rotation. In Landau gauge   $\bs{A}(r)=2(m/\hbar)\Omega x \hat{y}$ which is more suitable for the experimental set-up in ref. \cite{synthetic}.

The quantity $\nu=\frac{ 2 \Omega}{\frac{h}{m}} =-\frac{1}{2 \pi} \oint d\bs{r} \cdot \nabla \varphi_{i,j}$
gives the number of circulation quanta
through a unit cell in the square lattice and is gauge invariant.
For the $\nu=\frac{p}{q}$ ($p$ and $q$ are co-prime) as the boson hops around a unit cell in the square lattice it acquires a non trivial phase factor $exp(-2\pi i \nu)$. To achieve a winding which is integer multiple of the $2 \pi$, the boson should therefore hop around $q$  such unit cell leading to the formation of a magnetic unit cell \cite{Zak}.
This in turn implies that if we denote the phase of the  bosonic wave function  by the direction of an arrow then as one  goes around such magnetic unit cell, the arrow will rotate $p$ times and the wavefunction will have $-p$ vorticity in a magnetic unit cell. The same thing will happen even if we start from some
other lattice than the square lattice as long as the number of flux quanta goes through the unit cell will remain $\nu$.
Thus it imposes a topological constraint which does not depend on the local features such as the lattice potential..

The ground state of the Hamiltonian (\ref{RBH}) can be found by variational minimization of
$\langle \Psi |\hat{H}|\Psi \rangle$ with a Gutzwiller wave function $|\Psi \rangle =
\prod_{i} \sum_{n} f_{n}^{i}|n_i \rangle$. The variational parameters $f_n^{i}$ are the amplitudes for the Fock state $|n_i \rangle$ with $n$ particles at site $i$. A detailed analysis of  such  variational mean field approaches that is generally used to study the many body states of the Bose-Hubbard hamiltonian is given in reference \cite{Buonsante}.
For two dimensional lattice and $0 < V < \frac{1}{4} $ and $t=0$, the system will go through an alternating sequence of DW phase with $n_0$ and $n_0-1$ particles at successive sites, followed by a MI phase with $n_0$ particles per site where $n_0=1,2,3,\cdots$.
As $t$ increases a SS phase appears before the DW  state makes transition to a uniform superfluid phase.
%In the supersolid phase a spatially modulated superfluid order develops that co-exists with the crystaline order phase coherence is restored.

\subsection{Phase boundary for the non rotating case}
The phase boundary of the DW phase can be determined analytically by obtaining the energy of the particle-hole type excitations using a reduced basis
variational ansatz for the Gutzwiller wave function near the phase boundary.
The DW phase consists of two  sublattices $A$ and $B$ ( Fig. \ref{fig_latt} (a)) having fixed $n_0$ and $n_0-1$ particles per site. Thus it is convenient to decompose
$|\Psi \rangle = (|\Psi^{A} \rangle)( |\Psi^{B} \rangle)$. Here $|\Psi^{A} \rangle = \prod_{i_A=1}^{N/2} |\psi^{i_A} \rangle$ with
$|\psi^{i_A} \rangle =\sum_{n} f_{n}^{i_A}|n_{i_A} \rangle$ with $f_{n}^{i_A}=\delta_{n,n_0}$. Similarly
$|\Psi^{B} \rangle = \prod_{i_B=1}^{N/2} |\psi^{i_B} \rangle$ with
$|\psi^{i_B} \rangle=\sum_{m} f_{m}^{i_B}|m_{i_B} \rangle$ with $f_{m}^{i_B}=\delta_{m,n_0-1}$. For the non rotating case of $\Omega=0$,
very close to the DW phase boundary  only the neighbouring Fock
states are populated \cite{Oktel, Mueller1}. Thus for all $i$ and $j$
\bea |\psi^{i_A} \rangle & = & f_{n-1}^{i_A}|n-1 \rangle+ f_{n}^{i_A} | n \rangle + f_{n+1}^{i_A}|n +1 \rangle, n=n_0 \nonumber \\
     |\psi^{i_B} \rangle & = & f_{m-1}^{i_B}|m-1 \rangle + f_{m}^{i_B} | m \rangle + f_{m+1}^{i_B}|m+1 \rangle, m=n_0-1 \nonumber \eea
We set $(f_{n-1}^{i_A},f_{n}^{i_A},f_{n+1}^{i_A})  = (\epsilon_{1A}, \sqrt{1 - \epsilon_{1A}^2-\epsilon_{2A}^2}, \epsilon_{2A}) $
and $(f_{m-1}^{i_B},f_{m}^{i_B},f_{m+1}^{i_B}) =  (\epsilon_{1B}, \sqrt{1 - \epsilon_{1B}^2-\epsilon_{2B}^2}, \epsilon_{2B})$
with variational parameters $\epsilon_{1A,1B}$,$\epsilon_{2A,2B}$  all are $ \ll 1$ to ensure the normalization condition of states $|\psi^{i_A} \rangle, |\psi^{i_B} \rangle$. Also for the brevity of the notation
we have written $n_{i_A}$ as $n$ and $m_{i_B}$ as $m$ in the above expressions.
Minimization of the energy  with respect to these four parameters gives four equations. Their non trivial solution demands
\begin{widetext}
\beq
\det \left| \begin{array}{cccc}
(n -1) -\mu + 4V m & 0 & 4t \sqrt{nm} & 4t\sqrt{n(m+1)}  \\
 0 & (n - \mu + 4V m)  & -4t \sqrt{m(n+1)} & -4t \sqrt{(m+1)(n+1)} \\
 4t \sqrt{nm} & 4t \sqrt{m(n+1)} & (m-1) - \mu + 4Vn & 0 \\
-4t \sqrt{n(m+1)} & -4t\sqrt{(n+1)(m+1)}  & 0  &  (m - \mu + 4Vn) \end{array} \right | = 0 \label{phcdw} \eeq
\end{widetext}
A particle ($^p$) or  hole ($^h$) like excitation from either site of $A$ and $B$  are respectively given by
$\varepsilon_{p}^{A}  =  n + 4Vm , \varepsilon_{h}^{A} = -[(n-1) + 4Vm],~
     \varepsilon_{p}^{B}  =  m + 4Vn , \varepsilon_{h}^{B}= -[(m-1) + 4Vn] $.
Defining $\tilde{\varepsilon}_{p,h}^{A,B}= \varepsilon_{p,h}^{A,B} \mp \mu$ Eq. (\ref{phcdw}) gives the relation \cite{Kovrizhin}
\beq \tilde{\varepsilon}_{p}^{A}\tilde{\varepsilon}_{p}^{B}\tilde{\varepsilon}_{h}^{A}\tilde{\varepsilon}_{h}^{B} - (4t)^2
\left[ (n+1)\tilde{\varepsilon}_{h}^{A}+ n \tilde{\varepsilon}_{p}^{A} \right]
\left[(m+1)\tilde{\varepsilon}_{h}^{B} + m \tilde{\varepsilon}_{p}^{B}\right] = 0 \label{phexcite} \eeq
The above equation determines the
chemical potential $\mu$ at each $t$ for a given strength $V$ of the nearest neighbour interaction and will give the phase boundary. To understand significance of this equation in a better way we compare it with the similar results obtained 
within the framework of other mean field approaches such as time dependent Gutzwiller mean field theory \cite{Kovrizhin, Buonsante} where also minimal perturbation around a perfect density wave state is considered in the Fock space basis. 
Now, when such  particle or hole like excitation is created over a perfect density wave state, they do not remain localized at a site, but moves around the lattice to create a Bloch wave to minimize their energy. The kinetic energy of such a Bloch wave  is given by 
$\epsilon(\bs{k})=2t(\cos k_x + \cos k_y)$,  where $k_x, k_y$ are the components of the Bloch wave vector. The excitation spectrum of such particle-hole like excitations with finite wavevector can be obtained within the time dependent Gutzwiller mean field theory \cite{Kovrizhin} 
as \beq \tilde{\varepsilon}_{p}^{A}\tilde{\varepsilon}_{p}^{B}\tilde{\varepsilon}_{h}^{A}\tilde{\varepsilon}_{h}^{B} - \epsilon(\bs{k})^2
\left[ (n+1)\tilde{\varepsilon}_{h}^{A}+ n \tilde{\varepsilon}_{p}^{A} \right]
\left[(m+1)\tilde{\varepsilon}_{h}^{B} + m \tilde{\varepsilon}_{p}^{B}\right] = 0 \nonumber \eeq
where $\bs{k}=\{k_x,k_y \}$. The  density wave boundary  can again be retrieved by taking zero wave vector limit, namely $k_x \rightarrow 0,k_y \rightarrow 0 , \epsilon(\bs{k})=4t$, which expectedly reproduces our result in Eq. (\ref{phexcite}). We again emphasize that all the above displayed relations are for two dimensional square lattice, but can be generalized in other dimensions. 

In the next subsection we shall extend the above treatment for the rotating
case  and will show that the limiting particle-hole excitation spectrum that determines such phase boundary in presence of the finite rotation ( or magnetic field) is actually the edge  of a Hofstadter butterfly like energy spectrum.

\subsection{Rotated case}
For the rotated case, $\Omega  \neq 0$, we have
%\begin{widetext}
\bea \langle \Psi | \hat{H}| \Psi \rangle & = &
-2t \text{Re}\sum_{\langle i_A, i_B
\rangle}[ e^{i \varphi_{i_A i_B}}\phi_{A}^{i_A*} \phi_B^{i_B}]  \nonumber \\
&  & \mbox{} + \frac{1}{2}[\sum_{i=1}^{i=N} \sum_{n_i}
(n_i^2-n_i)|f_{n}^{i}|^2] -\mu
\sum_{i=1}^{i=N} \sum_{n_i} n_{i}|f_n^i|^2 \nonumber \\
& & \mbox{}+ V\sum_{\langle i_A, i_B \rangle} (\sum_{n_{A}}n_{A}|f_n^{i_A}|^2)(\sum_{m_{B}}m_{B}|f_{m}^{i_B}|^2)
\label{meanfield}\eea
%\end{widetext}
The first, second and fourth term respectively gives the mean kinetic, on site and nearest neighbour energy and the summation over $i$ in
the second and third term includes both the sublattices. In all further description again for brevity $n_{A}$ and $m_{B}$ will be written as $n$ and $m$
The superfluid order parameter on two sublattices (Fig. \ref{fig_latt} (b)) are given by
$\phi_{A}^{i_A},\phi_{B}^{i_B}=\langle \hat{a}_{i_{A}} \rangle, \langle \hat{a}_{i_B} \rangle$ whereas the DW order parameter is given by
$(-1)^{i}[\langle n_{i} \rangle - \frac{1}{N} \langle \sum_{i} n_{i} \rangle]$ on any site $i$ on either sublattices.
Near the DW phase boundary again only the neighbouring Fock states will get occupied. The
corresponding variational parameters
$(f_{n-1}^{i_A},f_{n}^{i_A},f_{n+1}^{i_A})$ for $i_A$ sites are $[\lambda_1^{i_A}
\Delta \phi_{A}^{i_A \ast},\sqrt{1 - |\Delta\phi_A^{i_A}|^{2}(|\lambda_1^{i_A}|^2 +
|\lambda_2^{i_A}|^2)}
,\lambda_{2}^{i_A} \Delta\phi_{A}^{i_A}]$, and, for $i_B$ sites we write
$(f_{m-1}^{i_B},f_{m}^{i_B},f_{m+1}^{i_B})$ as $[\delta_1^{i_B}
\Delta\phi_{B}^{i_B \ast},\sqrt{1 - |\Delta\phi_B^{i_B}|^{2}(|\delta_1^{i_B}|^2 +
|\delta_2^{i_B}|^2)}
,\delta_{2}^{i_B} \Delta\phi_{B}^{i_B}] $.
%The variational parameters  $\lambda_{1,2}$ and $\delta_{1,2}$ can be fixed by energy minimization.
The superfluid order parameters on the two sublattices are respectively given by 
$\phi_{A}^{i_A}=\sum_n\sqrt{n+1}f_n^{i_A*}f_{n+1}^{i_A}$ and $\phi_{B}^{i_B}=\sum_m\sqrt{m+1}f_m^{i_B*}f_{m+1}^{i_B}$. 
From these definitions it can be shown
$\phi_A^{i_A}  =  \Delta \phi_{A}^{i_A}+\textsl{O}((\Delta \phi_A^{i_A})^{3})$
with $\lambda_{2}^{i_A}=\frac{1}{\sqrt{n+1}}(1-\sqrt{n}\lambda_1^{i_A})$ and similarly
$\phi_B^{i_B}  =  \Delta \phi_{B}^{i_B}+\textsl{O}((\Delta \phi_B^{i_B})^{3})$ with
$\delta_{2}^{i_B} =  \frac{1}{\sqrt{m+1}}(1-\sqrt{m}\delta_{1}^{i_B})$
Thus if we neglect third and higher order corrections, $\Delta \phi_{A,B}$ can be replaced by the superfluid order parameter
$\phi_{A,B}$ on the two sublattices. Substituting these replacements 
and the expressions for variational parameters in the Eq. \ref{meanfield}
we obtain 

\begin{widetext}
\bea \langle \Psi | \hat{H}| \Psi \rangle & = &  -2t \text{Re}\sum_{\langle i_A, i_B
\rangle}[ e^{i \varphi_{i_A i_B}}\phi_{A}^{i_A*} \phi_B^{i_B}]
+\sum_{i_A}\left[{\frac{(n-\mu+4Vm)}{n+1}
\left[ 1-2\sqrt{n}|\lambda_1^{i_A}|-|\lambda_1^{i_A}|^{2}\right]+
|\lambda_1^{i_A}|^{2}}\right]|\phi_A^{i_A}|^{2}\nonumber\\
& & \mbox{} + \sum_{i_B}\left[{\frac{(m-\mu+4Vn)}{m+1}
\left[ 1-2\sqrt{m}|\delta _1^{i_B}|-|\delta_1^{i_B}|^{2}\right]+
|\delta_1^{i_B}|^{2}}\right]|\phi_B^{i_B}|^{2}
+E_{G} \label{ERBH2
} \eea
\end{widetext}
with $E_G$ is the energy of the pure density wave state.
\begin{figure}[ht]
\centerline{ \epsfxsize 10cm \epsfysize 7cm \epsffile{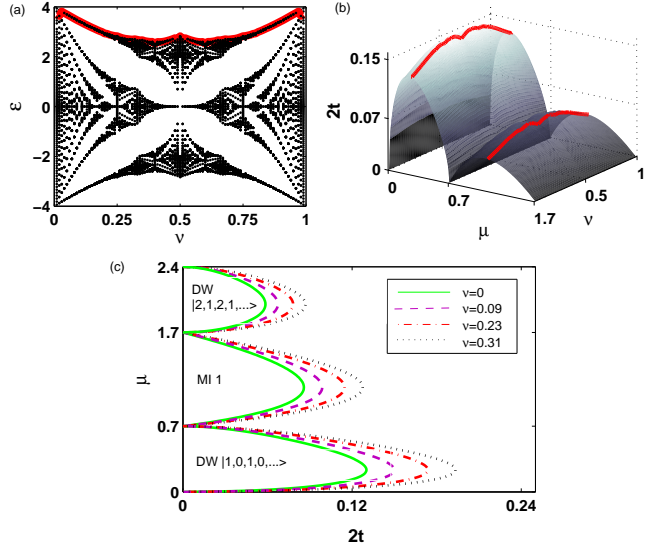}}
\caption{{(a) Hofstadter butterfly: the energy ($\varepsilon$) spectrum for the Eq. (\ref{harpereq}) for various $\nu \{0,1\}$. The upper
edge (marked red) gives the boundary of the density wave and the Mott Insulator lobe as explained in the text.
(b) The first DW and MI lobe as a function of $t, \mu, \nu$ in mean field approximation where $t$ and $\mu$ are in the unit of $U$.
(c) Cross section of the plots in (b) that shows the modification of the first two density wave lobe
and the first Mott lobe at various values of circulation quanta $\nu$. In all these plots $V$ has been taken as $0.2$ in the
unit of $U$.
}}
\label{fig1}
\end{figure}
To obtain the ground state energy, $\langle \Psi | \hat{H}| \Psi \rangle$  is first minimized with respect to  $\lambda_1^{i_A}$ and $\delta_1^{i_B}$
yielding $\lambda_1^{i_A}  = \sqrt{n} \frac{n - \mu + 4Vm}{1 + \mu -4Vm} ; \delta_1^{i_B}  =  \sqrt{m} \frac{m - \mu + 4Vn}{1 + \mu -4Vn}$.
Substituting the above expressions in $\langle \Psi | \hat{H}| \Psi \rangle$
, and, setting $\tilde{\phi_{A}^{i_A}}=\sqrt{\epsilon_1}\phi_{A}^{i_A}$ and $\tilde{\phi_{B}^{i_B}}=\sqrt{\epsilon_2}\phi_{B}^{i_B}$  and $\tilde{t}=\frac{t}{\sqrt{\epsilon_1 \epsilon_2}}$
where $\epsilon_1 = \frac{(n-\mu+4Vm)}{n+1} \left[ 1- n\frac{n - \mu + 4Vm}{1 + \mu -4Vm} \right]$ and $\epsilon_{2}=\epsilon_{1}(m \rightarrow n,n \rightarrow m)$, gives us the energy functional $\mathcal{E}$ near the DW phase boundary as
\bea \mathcal{E} & = & -\tilde{t}\sum_{\langle i_A,i_B \rangle} \begin{bmatrix} \tilde{\phi}_A^{i_A*} & \tilde{\phi}_B^{i_B*} \end{bmatrix}( \hat{\bs{n}} \cdot \bs{\sigma}) \begin{bmatrix} \tilde{\phi}_A^{i_A} & \tilde{\phi}_B^{i_B} \end{bmatrix}^{T} \nonumber \\
             &   & \mbox{} + \sum_{i_A} |\tilde{\phi}_A^{i_A}|^2 + \sum_{i_B}|\tilde{\phi}_B^{i_B}|^2 +E_{G}\label{DWLG} \eea
%            \text{with}~\epsilon_1 & = &  \frac{(n-\mu+2dVm)}{n+1} \left[ 1- n\frac{n - \mu + 2dVm}{1 + \mu -2dVm} \right] \nonumber \\
%             \epsilon_2 & = &  \frac{(m-\mu+2dVn)}{m+1} \left[ 1- m\frac{m - \mu + 2dVn}{1 + \mu -2dVn} \right] \nonumber \eea
The unit vector $\hat{\bs{n}} = \cos \varphi_{i_A i_B} \hat{x} + \sin \varphi_{i_A i_B} \hat{y}$ and $\bs{\sigma} = \sigma_x \hat{x} + \sigma_y \hat{y}$, where $\sigma_{x,y}$ are the Pauli matrices. The reduced basis ansatz  assumes very low superfluid density ($\phi_{A,B} \ll 1$). This is valid
very close to the phase boundary. Thus $\mathcal{E}$ contain terms only
linear in the superfluid density. This is unlike the Gross-Pitaevskii energy functional which contains terms quadratic in the superfluid  density
and is valid deep inside the superfluid regime.

Minimization of the above energy functional with respect to $\tilde{\phi}_{A}^{i_A*}, \tilde{\phi}_{B}^{i_B*}$ gives equations for the superfluid order parameter that can be written as
a spinorial Harper equation,
\beq \sum_{\langle i_{A},i_{B} \rangle}(\hat{\bs{n}} \cdot \bs{\sigma})
\begin{bmatrix} \tilde{\phi}_{A}^{i_A} & \tilde{\phi}_{B}^{i_B} \end{bmatrix}^{T}=\frac{1}{\tilde{t}} \begin{bmatrix} \tilde{\phi}_{A}^{i_A} & \tilde{\phi}_{B}^{i_B} \end{bmatrix}^{T} \eeq
Its solution can be written as
$\tilde{\phi}(x,y) \otimes  \begin{bmatrix} \exp(-i \frac{\varphi_{i_A i_B}}{2}) & \exp(i \frac{\varphi_{i_A i_B}}{2}) \end{bmatrix}^{T}$ where $\tilde{\phi}(x,y)$ satisfies the following symmetric gauge
Harper equation \cite{Harper}
\bea \tilde{\phi}(x+1,y) e^{i \pi \nu y} + \tilde{\phi}(x-1,y) e^{-i\pi \nu y} &  & ~~~\nonumber \\
\mbox{} +\tilde{\phi}(x,y+1)e^{-
i \pi \nu x}+ \tilde{\phi}(x,y-1) e^{i \pi \nu x} &  = & \frac{1}{\tilde{t}} \tilde{\phi}(x,y) \label{harpereq} \eea
$\frac{1}{\tilde{t}}$ in the right hand side of the Eq. (\ref{harpereq}) can be mapped on
the eigenvalues $\varepsilon$ of
HB \cite{Hofstadter} spectrum plotted in Fig.\ref{fig1} (a).
\section{Results and discussion}
\begin{figure}[ht]
\centerline{ \epsfxsize 10cm \epsfysize 6cm \epsffile{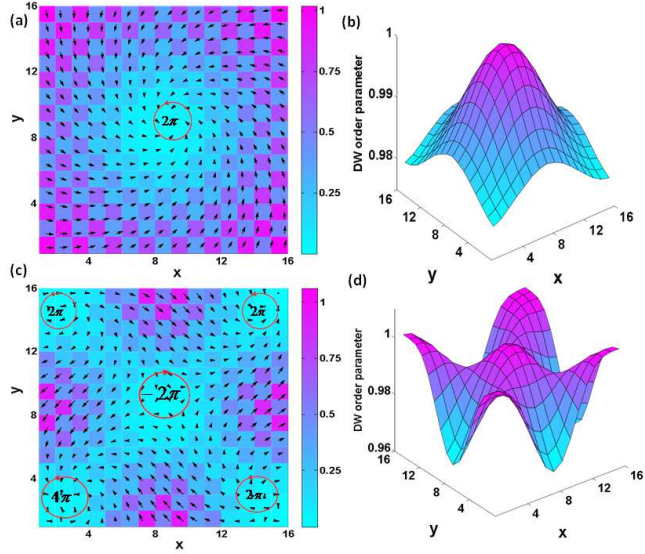}}
\caption{{(a) checkerboard vortices at the density wave  ( $|2,1,2,1,\cdots \rangle$) phase boundary ($\tilde{t}=\tilde{t}_c$) corrrsponding to the highest eigenvalue (the edge) of the
hofstadter butterfly spectrum for $\nu=\frac{1}{16 \times 16}$. The direction of the arrow gives $\varphi_{i_A,i_B}$
where as the color axis gives the superfluid density. The superfluid density is normalized by the maximum superfluid density at the boundary.
(b) corresponding DW order parameter (c) More complicated vortex structure corresponding to the higher value of $\tilde{t}$ corresponding to a lower eigenvalue ($
(254 (16 \times 16 -2)$th band)(
d) corresponding DW order parameter.
}}
\label{fig2}
\end{figure}

The edge of the HB spectrum (marked red in Fig. \ref{fig1} (a))
gives the highest eigenvalue of the Eq. (\ref{harpereq}) as function of $\nu \{0,1\}$. This corresponds to
the minimum value of $\tilde{t}=\tilde{t}_c=\frac{t_c}{\sqrt{\epsilon_1 \epsilon_2}}$ with non vanishing $\tilde{\phi}$ for each given value of $\mu$, and, hence
the boundary of the DW  phase at that particular $\nu$ ( marked red in Fig. \ref{fig1} (b)). Same observation holds true for
MI boundary at the MI-SF transition  in a rotating optical lattice \cite{Oktel, Mueller1}. Setting $m=n$ in the preceeding analysis
the MI-SF transition in rotated lattice can be studied for Extended Bose Hubbard model.  The phase boundary of the  ordinary Bose Hubbard model
under rotation or magnetic field can be retrieved by setting $V=0$ and also putting $n=m$ in the preceding analysis. The results obtained in this
way matches with those in reference \cite{Oktel, Mueller1}.  As compared to the modification of the phase boundary of
a MI phase in ordinary BH model here also the phase boundary of the DW phase extends as the strength of the gauge field
$\nu$ is enhanced. This is due to the stronger localization of the bosonic states by the  increasing strength of the gauge field. However
the superfluid order parameter of the excitations at the boundary of the DW phase are different from those near the MI boundary as we shall see in Fig.\ref{fig2}.
The modification of  DW as well as MI phase boundaries are plotted in Fig. \ref{fig1}(b) and (c). However it is important to note that the Fig. \ref{fig1}(c) only provides analytically the phase boundary of DW as well as MI phases within reduced basis ansatz and does not provide the phases themselves over the entire $t-\mu$ plane for various $\nu$  unlike in the references \cite{ Kovrizhin, Bartouni, Pai}. Particularly in mean field approximation it can be obtained numerically by using the full Gutzwiller wavefunction as was done in \cite{Kovrizhin}. 

%Hofstadter butterfly spectrum
%\cite{Hofstadter} originally describes the band structure of  electrons hopping on a lattice in presence of magnetic flux. The
%Hall conductivity of such electrons are related to topologically invariant Chern numbers \cite{topological} and are robustly
%quantized. The Chern number can also be calculated from the Hofstadter butterfly spectrum obtained in the present problem. However first of all %the current problem is being concerned about neutral bosonic atoms in optical lattice,  one cannot directly relate the Chern number described %above with a physically realizable quantity. Moreover
%here the DW or MI phase boundary of neutral bosons hopping on a lattice in presence of circulation quanta
%$\nu=\frac{p}{q}$ is obtained from the edge of the HB spectrum.  However, in  a recent work \cite{Goldman} it has been shown that this edge
%is also related with a topologically invariant Chern number $C_s$ through a Diophantine equation of
%the form $C_s = \frac{s_1}{\nu} - \frac{1}{p}$ where $s_1$ is an integer and $C_s \le \frac{q}{2}$. The preceeding discussion thus signifies the %topological nature of such phase boundary and implies that  they will remain same irrespective of the local features such as the nature of the %optical lattice potential etc.

At $\tilde{t}=\tilde{t}_c$ and $\nu=\frac{1}{L^2}$
each magnetic unit cell that consists of $L \times L$ lattice sites, contains one single vortex of unit winding. The strong sublattice modulation of
the superfluid density around the vortex core is shown in Fig. \ref{fig2}(a) for $L=16$.
The  DW order parameter given in Fig. \ref{fig2}(b)  becomes $1$ at the vortex core and co exists alongside the superfluid order in the bulk. Since at $\tilde{t}=\tilde{t}_c$ the systems become a supersolid in the mean-field Gutzwiller approximation
\cite{Kovrizhin}, the vortex structure in Fig. \ref{fig2}(a) correspond to the vortex structure just at this transition boundary.

We know that in a Hofstadter butterfly problem, for $\nu=\frac{p}{q}$, a given degenerate Landau level
is broken into $q$ bands for $p$ fluxes through a given magnetic unit cell. In our present case we have taken $\nu=\frac{1}{256}$.
The highest of these energy levels correspond to the critical value of $\tilde{t}=\tilde{t}_c$ at the phase boundary. Thus one may think that the eigenfunction for the lower energy levels that correspond to higher values of $\tilde{t}$ can be related with the superfluid phases away from the phase boundary of the DW state inside the supersolid regime. However this simplistic argument is not completely correct since the entire derivation presented above is only within the reduced basis ansatz, which is valid for $\tilde{t} \sim \tilde{t}_c$. Nevertheless we also plot the eigenfunction corresponding to a band which is very close to the highest band in  in Fig. \ref{fig2}(c). This approximately  depicts the superfluid order parameter in a rotated
supersolid phase for $\tilde{t} >
\tilde{t}_c$ , but still very close to the DW phase boundary.  This state, corresponding to the lower band of the same spectrum contains multiple
vortices in a given magnetic unit cell and the winding number of these vortices could also be integers $>1$. Such a vortex structure is plotted in Fig. \ref{fig2}(c). For calculating vortex structure at higher $\tilde{t}$ that corresponds to deep inside the supersolid phase, one needs to go beyond the reduced basis ansatz and includes the non linear terms due to superfluid interaction.

Experimental detection of such vortices near the phase boundary is possible with the presently available techniques. The sublattice modulation
of the superfluid density can be detected through the time of flight measurement and studying the resulting interference pattern \cite{Greiner}. To measure the detailed vortex structure in a magnetic unit cell
one can use Bragg scattering technique \cite{Bragg} which is sensitive to the spatial phase distribution of the initial state \cite{Blakie},
direction of rotation \cite{Chandra} and thus provides us a robust signature of the vortex state.

To conclude we showed that the modification of phase boundaries of an EBH model due to rotation induced artificial magnetic field can be
derived from the edge spectrum of a spinorial Harper equation.
From the spectrum of the same equation we have explicitly demonstrated within mean field theory how the superfluid and crystal order co-exist in the
vortex profile of a supersolid around a DW vortex core.
 This can be used to identify the exotic supersolid phase in cold atom experiments.
The above calculation can be
generalized for the other variants of the EBH model such as one that includes next nearest neighbour interaction and can motivate further study
for such vortices by going beyond the mean field approximation.

We thank G. V. Pai, D. Goldbaum, E. Mueller, K. Seshadri, J. Avron, O. Gat and E. Altman, S. Sinha for helpful correspondences. The work of RS is supported by CSIR, India and the work of SG is supported by the Planning unit of IIT Delhi.

\end{document}